Original Paper

Goran Murić ,[1] Kristina Lerman [1,2] and Emilio Ferrara [1,2,3]
1. USC Information Sciences Institute
2. USC Department of Computer Science
3. USC Annenberg School for Communication and Journalism

# Gender disparity in the authorship of biomedical research publications during the COVID-19 pandemic

## Abstract

**Background:** Gender imbalances in academia have been evident historically and persist today. For the past 60 years, we have witnessed the increase of participation of women in biomedical disciplines, showing that the gender gap is shrinking. However, preliminary evidence suggests that women, including female researchers, are disproportionately affected by the COVID-19 pandemic in terms of unequal distribution of childcare, elderly care and other kinds of domestic and emotional labor. Sudden lockdowns and abrupt shifts in daily routines have disproportionate consequences on their productivity, which is reflected by a sudden drop in research output in biomedical research, consequently affecting the number of female authors of scientific publications.

**Objective:** The objective of this study is to test the hypothesis that the COVID-19 pandemic has a disproportionate adverse effect on the productivity of female researchers in biomedical field in terms of authorship of scientific publications.

**Methods:** This is a retrospective observational bibliometric study. We investigate the proportion of male and female researchers who published scientific papers during the COVID-19 pandemic, using bibliometric data from biomedical preprint servers and selected Springer-Nature journals. We use the Ordinary Least Squares (OLS) regression model to estimate the expected proportions over time by correcting for temporal trends. We also use a set of statistical methods such as Kolmogorov-Smirnov (KS) test and Regression Discontinuity Design (RDD) to test the validity of results.

**Results:** A total of 78,950 papers from bioRxiv, medRxiv and 62 selected Springer-Nature journals by 346,354 unique authors were analyzed. The acquired dataset consisted of papers that were published between January 1, 2019, and August 2, 2020. The proportion of women first authors publishing in biomedical field during the pandemic drops on average 9.1% across disciplines (expected arithmetic mean $\bar{y}_{est} = 0.39$; observed arithmetic mean $\bar{y} = 0.35$; standard error of the estimate, $\bar{S}_{est} = 0.007$; standard error of the observation, $\sigma_{\bar{x}} = 0.004$). The impact is particularly pronounced for papers related to COVID-19 research, where the proportion of female scientists in the first author position drops by 28% ($\bar{y}_{est} = $ 0.39; $\bar{y} = 0.28; \bar{S}_{est} = 0.007; \sigma_{\bar{x}} = 0.007$). When looking at the last authors, the proportion of women drops in average 7.9% ($\bar{y}_{est} = 0.25; \bar{y} = 0.23; \bar{S}_{est} = 0.005; \sigma_{\bar{x}} = 0.003$), while the proportion of women writing about COVID-19 as the last author decreased by 18.8% ($\bar{y}_{est} = 0.25; \bar{y} = 0.21; \bar{S}_{est} = 0.005; \sigma_{\bar{x}} = 0.007$). Further, by geocoding authors' affiliations, we show that the gender disparities become even more apparent when disaggregated by the country, up to 35% in some cases.

**Conclusions:** Our findings document a decrease in the number of publications by female authors in biomedical field during the global pandemic. This effect is particularly pronounced for papers related to COVID-19, indicating that women are producing fewer publications related to COVID-19 research. This sudden increase in the gender gap is persistent across the ten countries with the highest number of researchers. These results should be used to inform the scientific community of the worrying trend in COVID-19 research and the disproportionate effect that the pandemic has on female academics.

## Introduction

As of the date of this writing, the COVID-19 pandemic has claimed hundreds of thousands of lives worldwide and disrupted almost all aspects of human society. The socio-economic impacts of the pandemic are yet to be assessed and the impending economic crisis and recession are becoming evident.[1–3] During the recessions, men are more likely to lose their job as men work in industries which are heavily affected by the slowdown in economic activity such as manufacturing and construction. Compared to previous economic crises, the current crisis has disproportionately affected female workers. [4–11] One of the reasons for such disparity is women's over-representation in occupations in industries that are most affected by the closures and movement restrictions, such as restaurants and hospitality. Another large part of gender disparity is related to the unequal division of labor in the household, as women are traditionally expected to continue to devote more time to child care and domestic chores than their partners. [4] In case of the dual-earner, heterosexual married couples with children the partners have unequally adjusted their work time during the pandemic. Mothers with young children have reduced their work hours four to five times more than fathers contributing to the increased gender gap in earnings. [5] Working mothers affected by the unequal distribution of working hours and the additional burden of domestic chores have reported lower work productivity and job satisfaction than men. [6]

Stay-at-home orders, lockdowns and school closures have affected scientists as well, especially those caring for children or other family members. [12,13] Female scientists reported that their ability to devote time to their research has been substantially affected, and the impact is most pronounced for female scientists with young dependents. [14] The sudden shift in daily activities makes it hard to balance between increasing professional requirements and childcare.

As a result, the research productivity of female scientists appears to have decreased. [15–18] Early evidence suggests that the proportion of publications with female authors is lower during the pandemic with the evident gendered authorship disparities in journal submissions. [19,20] Reports from journal editors in the fields of International Studies, Political Science, Economics, Medicine, and Philosophy indicate that the proportion of submissions authored by women, in most cases has dropped. [21] Even though while female academics are still submitting manuscripts for publication during the crisis, they are submitting less of their own work than men. [22]

A similar effect has been observed with publications on preprint servers. The proportion of female authors publishing on the most popular economics preprint servers is lower than expected, [23,24] with only 14.6 percent of female authors, whereas comparably, they usually make up about 20 percent of authors in these databases. Similarly, women publish less in other disciplines such as physics, earth science and sociology. [25] In regards to medical and related sciences, on top of the exacerbated gender disparity in publishing during the pandemic, the proportion of female scientists publishing research specifically about COVID-19 is much lower than expected, by almost 23%. [25–27]

Motivated by the ongoing research efforts, we expand on the previous research by analyzing the large bibliographic dataset in biomedical field; we also employ different modelling techniques that can further improve our understanding of this phenomenon. The aim of this study is to quantify how COVID-19 exacerbates the gender gap in scientific publishing in biomedical field.

## Materials and Methods

### Data

The bibliometric data on published papers are collected from three separate sources: 1) *bioRxiv* (51,171 papers and 225,110 authors), provided by the Rxivist, the API provider for bioRxiv publications; [28] 2) *medRxiv* (8,845 papers and 52,364 authors), scraped directly from medrxiv.org; 3) *Springer-Nature* (19,525 papers and 91,257 authors) data from 62 journals collected using the *Springer-Nature OpenAccess API*. *Springer-Nature* data includes high-impact journals such as *Nature Genetics*, *Nature Medicine* and *Nature Immunology,* as well as multiple *BMC* journals such as *BMC Bioinformatics* and *BMC Genomics*. We include the data from all journals in biomedical field that Springer-Nature provides data for. A complete list of journals used in the analysis is available in Multimedia Appendix 1, Table S7. All the papers from the dataset were published between January 1, 2019, and August 2, 2020. The earliest publication on *medRxiv* is from June 25, 2019.

For each source, we collect the relevant meta-data. For each paper in *bioRxiv* and *medRxiv*, we keep the *date* of publishing. For *Springer-Nature* journals, we keep the date of manuscript submission, which is the most comparable date to publication dates in *bioRxiv* and *medRxiv*. We additionally store the *title* and the *abstract* of the papers as well as the scientific discipline of the paper. There is a total of 112 scientific disciplines in the data, and each paper belongs to a single discipline. The complete list of all disciplines is provided in the Multimedia Appendix 1, Table S11. For each author, we preserve the *name*, *affiliation* and the authorship order. We remove the papers with group authors, such as scientific consortiums and projects ($\sim 0.1\%$ of all papers), as they do not represent individuals. We use socio-economic data on countries, including their respective GDP per capita provided by *Our World in Data* (ourworldindata.org).

### Identifying authors' gender

To infer each author's gender from their name, we use a state-of-the-art tool, namely the *genderize.io API.* [29] Given an input name, the model returns a gender and a confidence score between 0.5 and 1. The uncertainty is greater for Asian names that often are not gender-specific. [30] We filter out all authors for which the confidence score is lower than 0.8. Overall, 19% of names yields a score below this threshold, with Chinese and Korean topping the ranking with 54% and 41%,' respectively.
In our dataset, we identify the most likely gender of the 466,836 authors in total. Out of these, the gender of 348,506 unique authors (214,095 male and 134,411 female) could be inferred with high accuracy with a *genderize's* confidence score higher than 0.8, which is 74% of all authors.

### Identifying authors' country
To identify each author's country in the *bioRxiv* and *medRxiv* dataset, we first locate a toponym in the author's affiliation and assign the most likely country code to a given toponym. If there is no toponym, we query the GRID.ac database and find the institution with the most similar name and assign the institution's country to the author. Additionally, we manually check the location of the most common affiliation names from the dataset that covers most of the authors. The country of approximately 80% of all authors was determined using this method. The country of the authors in *Springer-Nature* dataset was already provided by the API.

### Identifying COVID-19 papers

The papers that deal specifically with COVID-19 and similar topics are identified by the set of keywords that appear in the title or in the abstract.

### Calculating the differences between the expected and observed proportions

To measure the discrepancy between the expected and the observed proportion of female authors, we first establish the baselines, which are the expected proportions of female researchers that appear as the authors of publications. The expected proportions are calculated using the OLS model and historical data from January 2019 to March 2020 (see Model). We then calculate the true observed proportions of female authors who published during the COVID-19 pandemic in 2020 and compare it to the expected baselines. The error for the predicted value is the mean standard error of the prediction. The error of the observed value is calculated as the standard error of the mean $SE = \sigma/\sqrt{n}$. The percentage change is calculated as $diff = (f^{exp} - f^{obs})/f^{exp}$, where the $f^{exp}$ is the expected proportion and $f^{obs}$ is the observed proportion. The errors for the percentage change in Figure 3 are calculated as the total sum of the errors of predicted and observed values.

### Model

Using historical data before March 15, 2020, we calculate the proportion of female authors publishing each week. We fit an ordinary least squares (OLS) regression model $f = \beta t + c$, where $f$ is the proportion of female authors that serves as a response variable, $t$ is the predictor variable - time of publication/submission (to the nearest week), $\beta$ and $c$ are the slope and the intercept. We fit the separate models depending on the level of disaggregation (country, publisher, ...). The model is illustrated in Figure 1. From the model, we derive the expected fraction $f^{Exp} = \sum \hat{f}/n$ that is the mean fraction of all predicted values for the observed period and $f^{Obs} = \sum f_{true}/n$. To estimate the expected number of papers and authors, we use a similar approach, where the response variables are the numbers of papers and authors rather than the proportion of female authors. We use *statsmodels* [31] package in *Python* for this purpose.

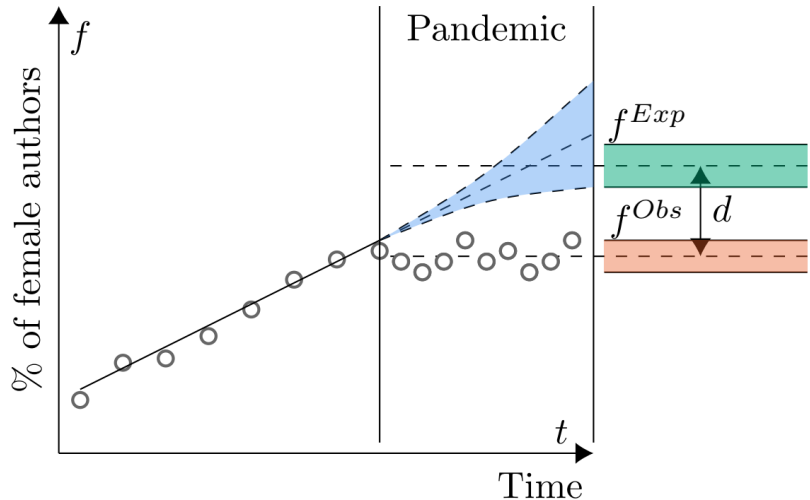

*Figure 1.* **Statistical model.** Schematic illustration of the OLS model used to calculate the expected numbers and proportions.

The OLS model tends to weight equally all data points regardless of the number of samples. To guarantee the validity of the statistical analysis, we establish the conditions under which the data points will be evaluated. The number of data points used to fit the OLS before March 2020 and the number of data points after March 2020 are at least 10 each. This way, we limit the impact of small-sample observations that can skew the estimate.

We additionally evaluate the model by applying the generalized linear model with binomial errors and a logit link function, as the OLS could overestimate the proportions in binary variables. Both models perform similarly, and the OLS model did not provide any out-of-norm estimates. For the sake of better interpretability and consistency with modeling the nominal number of authors and papers, we decide to use the OLS model.

To better capture the productivity of the population, we count each publication of the author separately, effectively modeling the proportion of papers authored by the population of female authors. Considering that multiple authorships in the observed period are relatively rare (less than 5% of all first authors and less than 10% of all last authors have more than one paper), we consider each authorship independently.

### Regression discontinuity design

To estimate the potential causal effects of the pandemic on the proportion of female researchers, we devise a typical non-parametric regression discontinuity design with a local linear regression in time, with the general form:

$$Y = \alpha + \tau D + \beta_1(X - c) + \beta_2 D(X - c) + \epsilon$$

where $c$ is the treatment cutoff and $D$ is a binary variable equal to one if $X \geq c$. In our case, we assume the date $c$ of a policy change that is mid-March 2020. For all dates $t > c$, the unit is treated, and for all dates $t < c$, the unit is not. This regression discontinuity setup uses time-series data, and in this case, weekly observations, both globally and on a country level. By comparing observations lying closely on either side of the temporal threshold, we estimate the average treatment effect. We make sure to focus on observations not too far (in time) from the threshold, avoiding potential bias from unobservable confounders. [32]

The falsification (or placebo) tests are performed by using fake cutoffs before and after mid-March 2020 and comparing the treatment effect. We identify the optimal cutoff point $c_0$ as the point in time when the treatment effect is the most prominent, $c_0 = c|\max(|\tau|)$. The regression discontinuity design has been implemented using the *rdd* package in *Python.* [33]

### Data availability and reproducibility
The data and the source code for reproducing the results are available at:
https://github.com/gmuric/CovidGenderGap

## Results

### The gender gap in research during the COVID-19 pandemic

Overall, during the pandemic, scientists posted papers on preprint servers at an increasing rate. On average, we observe 31.2% more papers than expected and a 41.6% increase in the number of authors (39.2% increase for females and 42.9% increase for males). Despite the absolute increase in the numbers of papers and authors across publishers (see Multimedia Appendix 1, Figure S1, Tables S1, S2 and S3), the proportion of female authors is lower than expected.

In biology, medicine and related disciplines, the most active contributors are usually listed first. The author listed last is the most senior author, typically the head of the lab. To address the high variability of the number of authors on the publications ($\mu = 7.4, \sigma = 9.2$), we analyze separately the proportion of women who appear as the first author, the last author, an author regardless of the authorship order and as the solo author. Additionally, we perform a separate analysis on the papers with topics that are related directly to COVID-19.

Table 1. The expected and observed proportion of female authors disaggregated by the order of authorship and the topic

| Order | Papers | Expected $\bar{y}_{est}$ | $\bar{S}_{est}$ | Observed $\bar{y}$ | $\sigma_{\bar{x}}$ | % drop |
|---|---|---|---|---|---|---|
| First | All | 0.389 | 0.007 | 0.353 | 0.004 | 9.142 |
|  | COVID-19 | 0.389 | 0.007 | 0.28 | 0.007 | 28.031 |
|  | Non COVID-19 | 0.389 | 0.007 | 0.38 | 0.004 | 2.372 |
| Last | All | 0.257 | 0.005 | 0.236 | 0.003 | 7.961 |
|  | COVID-19 | 0.257 | 0.005 | 0.209 | 0.007 | 18.812 |
|  | Non COVID-19 | 0.257 | 0.005 | 0.246 | 0.003 | 4.416 |
| Any | All | 0.354 | 0.003 | 0.348 | 0.002 | 1.578 |
|  | COVID-19 | 0.354 | 0.003 | 0.341 | 0.009 | 3.53 |
|  | Non COVID-19 | 0.354 | 0.003 | 0.351 | 0.002 | 0.934 |
| Solo | All | 0.21 | 0.03 | 0.137 | 0.008 | 34.586 |
|  | COVID-19 | 0.21 | 0.03 | 0.137 | 0.023 | 34.514 |
|  | Non COVID-19 | 0.21 | 0.03 | 0.168 | 0.013 | 19.802 |

$\bar{y}_{est}$ is the arithmetic mean of the estimate, $\bar{y}$ is the arithmetic mean of the observation
$\bar{S}_{est}$ is the mean standard error of the estimate
$\sigma_{\bar{x}}$ is the standard error of the mean (SEM) of the observation

The aggregate results suggest that the proportion of female authors publishing on all topics as the first author decreased by 9.1% (expected arithmetic mean $\bar{y}_{est} = 0.38$; observed arithmetic mean $\bar{y} = 0.35$; standard error of the estimate, $\bar{S}_{est} = 0.007$; standard error of the observation, $\sigma_{\bar{x}} = 0.004$). The percentage drop becomes unusually prominent when we analyze the papers about COVID-19. The proportion of female scientists who write on COVID-19 related topics as the first author is lower by 28% ($\bar{y}_{est} = 0.38$; $\bar{y} = 0.27$; $\bar{S}_{est} = 0.007$; $\sigma_{\bar{x}} = 0.007$). When considering the last authors, the proportion of women writing about COVID-19 decreased by 18.8% ($\bar{y}_{est} = 0.25$; $\bar{y} = 0.2$; $\bar{S}_{est} = 0.005$; $\sigma_{\bar{x}} = 0.007$). However, if we focus only on papers that do not deal with COVID-19, we see a smaller change both for the first author (2.3%) and the last author (4.4%). The proportion of women publishing papers on topics other than COVID-19 on *medRxiv* increased on average 14%. The results are shown in Table 1 and illustrated in Figure 2. The expected proportions are plotted as the green bars, and the true proportions are plotted in orange. The standard errors of the aggregate analysis are represented as the vertical lines on top of the bars and are relatively small (see Materials and Methods for details about the errors calculation).

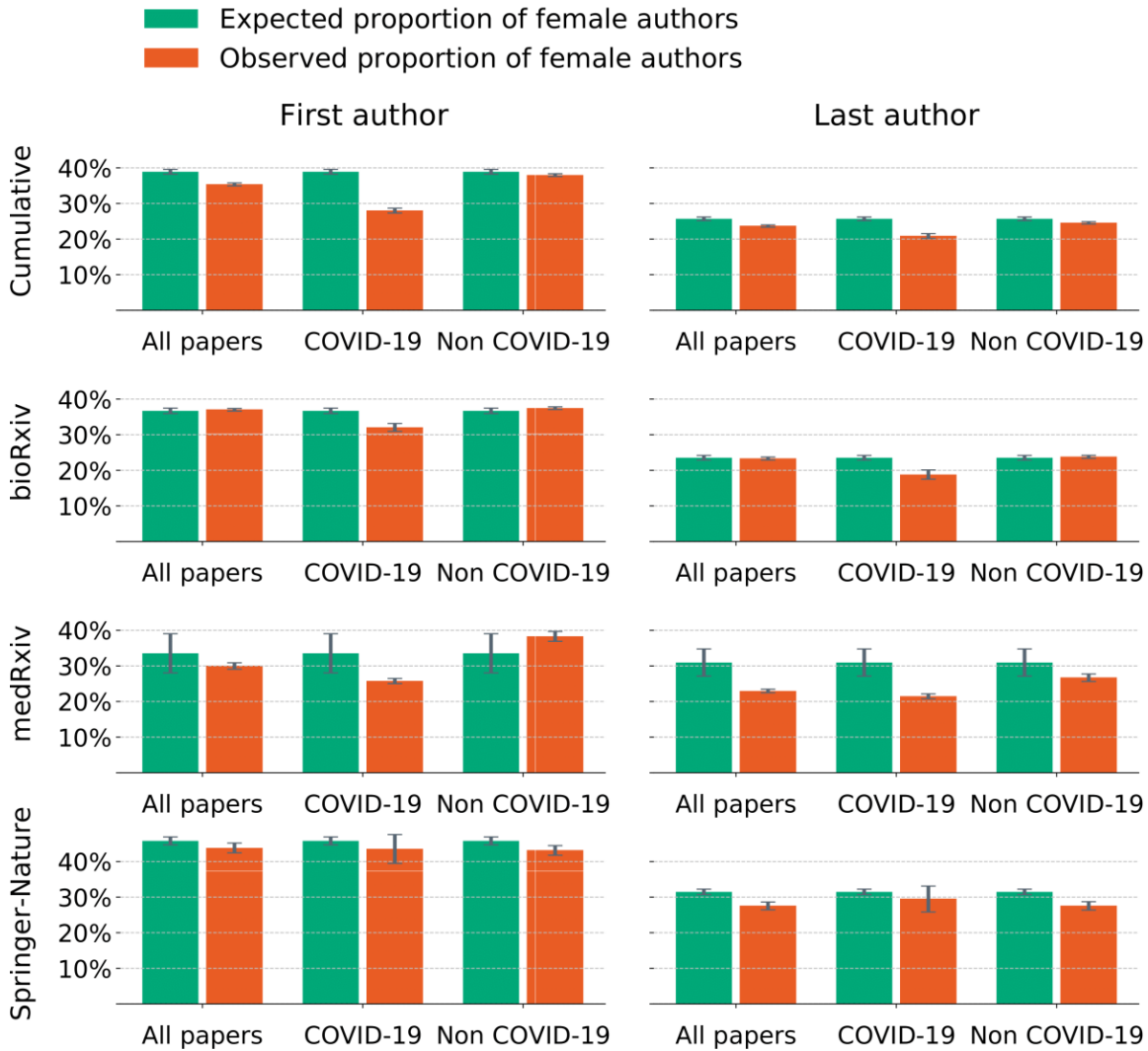

*Figure 2.* **The comparison of the expected and observed proportion of female authors who publish during the COVID-19 pandemic. Green bars represent the expected proportion of female authors, estimated by the OLS model from the historical data from 2019. Orange bars are the observed proportion of female authors that publish during the COVID-19 pandemic. The papers are divided by the topic in three groups: 1) all papers from the dataset; 2) the papers that deal directly with the COVID-19 and related topics; 3) the papers that are not about COVID-19 or related topics. In the first row are the results from all publishers combined. Other rows represent the results for each publisher separately.**

Additionally, we focus our analysis on the papers with a single author and discover even greater disparity. We observe 34.5% ($\bar{y}_{est} = 0.21$; $\bar{y} = 0.13$; $\bar{S}_{est} = 0.03$; $\sigma_{\bar{x}} = 0.008$) fewer female solo authors during the pandemic publishing on all topics across the platforms (see Multimedia Appendix 1, Figure S2). Similar disparity appears in case of the solo authors publishing papers about COVID-19. Note that only 3% of all papers are authored by a single author, hence the relatively large standard errors of the both estimate and the observation, especially for papers published on *medRxiv*. The effect still exists, although much less prominently, when we observe all authors regardless of the order of authorship. More detailed information is provided in Multimedia Appendix 1 in Table S1.

The results suggest that the aggregate gender disparity in academia during the pandemic is due to the increased publication rate of papers about COVID-19 authored by men. To further explore this possibility, we track the individual publication records and calculate the probability that the author will publish work about COVID-19. Around 3.7% of men who have publication records in our dataset will publish at least one paper about COVID-19, compared to $\approx$ 2.2% of women. Men who already have a publication before the pandemic are 37% more likely to publish a paper about COVID-19. This suggest that women are getting excluded from critical research about COVID-19.

When disaggregated by publisher, the relative drop in the proportion of female first authors for COVID-19 related research is 12.6%, 23.2% and 2.1% for *bioRxiv*, *medRxiv* and *Springer-Nature* journals respectively (Figure 2). Similar disparity is observed for the last authors with a relative drop of 20.1%, 30.8% and 23.6% and the authors regardless of the authorship order with a relative drop of 2.2%, 10.7% and 16.1%. In the case of solo papers, the average drop across the platforms is 34.5%. The proportion of females publishing on topics other than COVID-19 remains within the standard error of the estimate, without the strong evidence of decrease. Note the large standard errors in the estimated proportion of women for COVID-19 related papers in *Springer-Nature* journals due to the lack of data. Only published papers have meta-data available through the *Springer-Nature* API, and many papers submitted during the pandemic that will ultimately be published have not yet been accepted and published (see Materials and Methods).

Additionally, we check if there was a significant change in the proportion of women authors that occurred in mid-March 2020. To test the hypothesis, we perform a regression discontinuity design analysis in time (see Materials and Methods). We estimate a vertical discontinuity of the proportion of women over time by the coefficient $\tau$ at the cutoff point $c_0$ = March 15, 2020. For all the papers, regardless of the topic, we get $\tau = -0.008$ with $p = 0.03$. However, considering only the papers about COVID-19, the discontinuity becomes clearer with $\tau = -0.049$ and $p < 0.005$. To assert the robustness of our model, we perform a placebo test (see Materials and Methods) to confirm that the discontinuity is likely aligned with the start of the pandemic and that it happened at or around March 2020, and not during any time in 2019. When RDD analysis is performed at the country level, we confirm that for most countries, the cutoff threshold falls between mid-March and mid-April 2020 (see Multimedia Appendix 1 Table S7). The RDD analysis suggests that there was a drop in the proportion of female authors in the beginning of April 2020 that was more significant than any other fluctuation that occurred in 2019 or after April 2020.

Further, we check if we can confidently use the proportion of women who published before the pandemic as the reference to estimate the proportion of women to publish papers specifically about COVID-19. A hypothesis is that before the pandemic, women can be less likely to be represented in the scientific disciplines that will produce COVID-19 research. To check that, we first perform a chi-squared test on the distribution of disciplines involved in COVID-19 research. We discover that some disciplines, such as infectious diseases, epidemiology, public and global health are over-represented ($p < 0.01$). Then, we test if the proportions of women in COVID-19 disciplines are significantly different from non-COVID-19 disciplines. By performing the Kolmogorov–Smirnov test, we compare the distributions of the proportion of women across two groups of disciplines and we obtain $p = 0.84$. We conclude that the two groups were sampled from populations with the same distributions, and we can be confident that we can use the data on the proportion of women before the pandemic to model the proportion of women that publish about COVID-19.

### Some trends during the pandemic

To assess the temporal trends during the pandemic, we build the linear model $f(t) = \alpha + \beta t + \epsilon$, where $f(t)$ is the proportion of female scientist, and $t$ is the time in weeks after mid March 2020. The regression coefficient $\beta$ is used to quantify the trend. We did not identify a significant change in the proportion of female first authors (Table 2). However, we observe a small but significant increase in the proportion of female scientists appearing as the last author ($\beta = 0.001, p < 0.01$) and an author regardless of authorship order ($\beta = 0.001, p < 0.01$). An even stronger positive trend is observed for the COVID-19 related research for the last author ($\beta = 0.003, p < 0.01$) and an author regardless of authorship order with $\beta = 0.005$ (Multimedia Appendix 1, Table S8).

*Table 2.* The parameters of the linear model of the proportion of female authors over time during the pandemic

|  | First author | | Last author | | All authors | | Solo author | |
| --- | --- | --- | --- | --- | --- | --- | --- | --- |
|  | β | p | β | p | β | p | β | p |
| All papers | -0.002 | 0.083 | 0.002 | 0.002 | 0.001 | 0.0 | 0.000 | 0.999 |
| COVID-19 papers | 0.001 | 0.278 | 0.003 | 0.0 | 0.005 | 0.0 | -0.005 | 0.119 |
| Non-COVID-19 papers | -0.002 | 0.024 | 0.001 | 0.025 | 0.000 | 0.169 | 0.004 | 0.132 |

$\beta$ - a regression coefficient

### Country-level Analysis

We identify the most likely country of authors based on their affiliation (see Materials and Methods) and measure the difference between the expected and observed proportion of female authors during the pandemic. Figure 3 shows the pandemic-related gender gap across the countries with the largest share of the authors. The values represent percentage differences between the expected and observed fraction of female authors publishing in *bioRxiv*, *medRxiv* and selected *Springer-Nature* journals between March and August 2020. Points to the left (orange) of the mid-line represent countries with less than expected fraction of female authors, and points to the right of the mid-line (in green) represent an increase in the fraction of female authors. The left plot is for all papers regardless of topic, the middle plot is only for COVID-19 related papers and the right plot is only for the papers that are not related to COVID-19. More detailed information is provided in Multimedia Appendix 1 in Tables S4-S6.

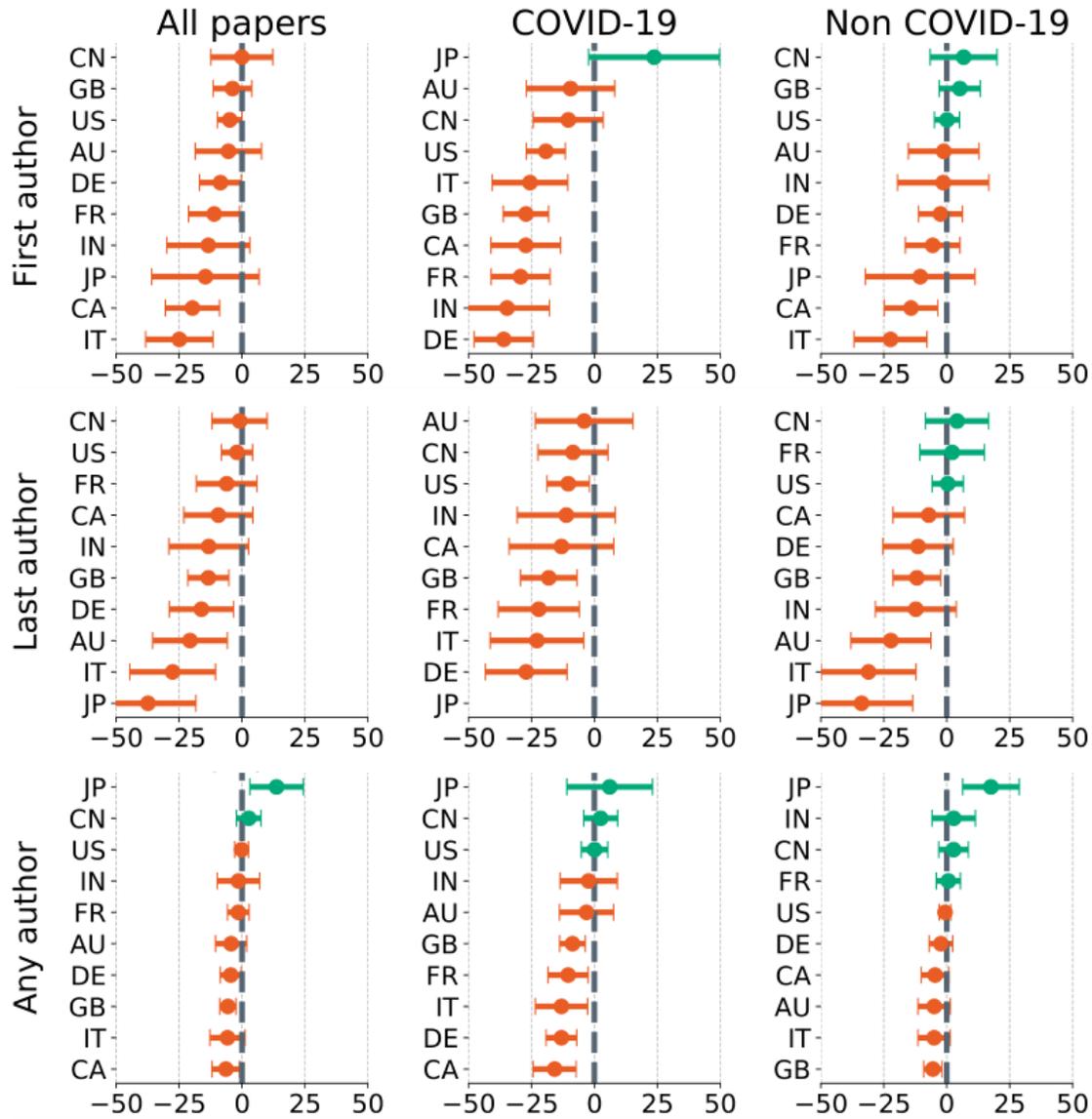

*Figure 3.* **Percentage drop in proportion of female authors during the pandemic across the countries.** Orange points mark the percentage decrease in proportion of female authors. Green points mark the increase. Horizontal lines are standard errors. The analysis is divided by the topic in three groups: 1) all papers from the dataset; 2) the papers that deal directly with the COVID-19 and related topics; 3) the papers that are not about COVID-19 or related topics. Missing points indicate insufficient sample size.

A significant drop in the proportion of female first authors is consistent across the countries. Regardless of the topic, we observe a 24.9% drop in Italy ($\bar{y}_{est} = 0.526$; $\bar{y} = 0.395$; $\bar{S}_{est} = 0.046$; $\sigma_{\bar{x}} = 0.024$), followed by Canada (19.7%), Sweden

(15.9%), Japan (14.5%), India (13.4%) and France (11.1%). For the research dealing explicitly with the topic of COVID-19 (Figure 3, middle panel), we observe a greater gender gap than with the papers on other research topics (Figure 3, right panel). In Germany, for example, the relative drop in the proportion of female first authors is 36% ($\bar{y}_{est} = 0.39$; $\bar{y} = 0.25$; $\bar{S}_{est} = 0.02$; $\sigma_{\bar{x}} = 0.027$), indicating that male scientists affiliated with German institutions are publishing disproportionately more than their female colleagues about COVID-19. Similar considerations apply to India, France, Italy, Great Britain, Canada and the United States. The opposite is true for Japan, where the proportion of women publishing about COVID-19 as the first author increases by 23.7%. Similar disparity applies to the last authors (Figure 3, second row). Missing points indicate that there was not enough data from the pandemic period to calculate the observed mean.

When we observe the proportion of female authors regardless of the authorship order, the drop becomes less prominent but still consistent across the countries. For example, in Canada, a drop in the proportion of female authors for COVID-19 related papers is 15.7% ($\bar{y}_{est} = 0.378$; $\bar{y} = 0.318$; $\bar{S}_{est} = 0.014$; $\sigma_{\bar{x}} = 0.018$), with similar-sized drops for countries such as Germany, Italy, Great Britain, and France (Figure 3, third row). On the other hand, there is an increase in the proportion of female authors writing about COVID-19 in Japan for 6% ($\bar{y}_{est} = 0.155$; $\bar{y} = 0.164$; $\bar{S}_{est} = 0.010$; $\sigma_{\bar{x}} = 0.017$). The increase in the proportion of female authors in China (2.4%) is within the margin of error.

The gender gap for non-COVID-19 related research (Figure 3, right panel) exists during the pandemic, but it is smaller than for COVID-19 research. Again, we observe the stark differences among countries, with the proportion of female first authors publishing during the pandemic significantly decreasing in Italy, Canada, Japan and France. Note that the plots for the single author papers are missing as the samples become too small when disaggregated by the country.

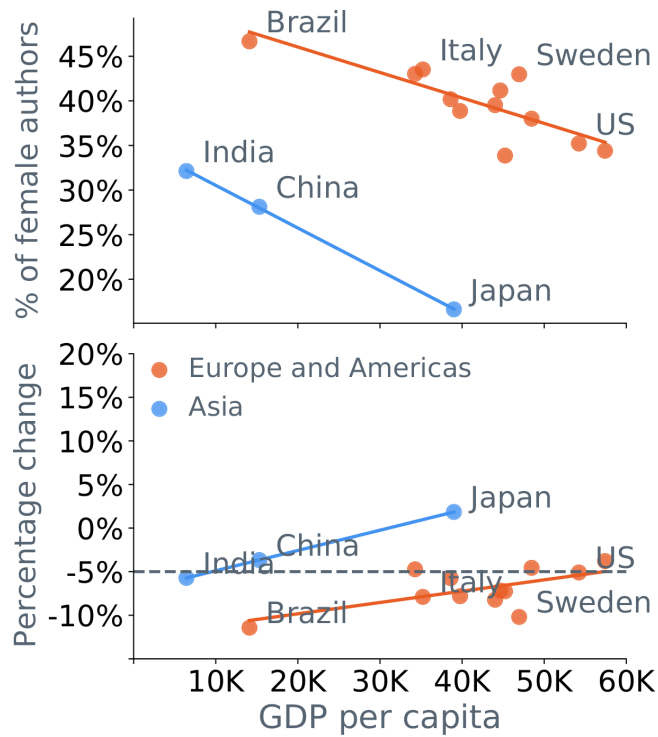

*Figure 4.* **Gender disparity in research and the GDP.** (Upper) The proportion of women active in research is higher in countries with lower per capita GDP. (Lower) The proportion of female authors of research articles decreased more than expected in countries with lower per capita GDP.

Further, we explore if there are any commonalities among the countries with respect to the participation of women in research. Figure 4 illustrates the proportions of female authors (upper panel) and the percentage change in the proportion of female authors (lower panel). When disaggregated by region, we observe that the wealthier countries—with higher per capita gross domestic product (GDP)—have proportionally fewer female researchers, with Asian countries exhibiting the most pronounced gender disparity. However, the countries with higher GDP per capita demonstrate a smaller drop in the proportion of women publishing during the pandemic.

## Discussion

We analyze bibliographical data from biomedical preprint servers and Springer-Nature journals and show that the fraction of women publishing during the COVID-19 pandemic drops significantly across disciplines and research topics. Since the announcement of the global pandemic and the start of lockdowns, we observe a drop of 9.1% in the number of women publishing biomedical scientific papers as the first author. Women are significantly excluded from COVID-19 related research, as we measure the 28% drop in female first authors in that area of research. That confirms some earlier suggestions that female first authors contribute less to COVID-19 studies than to research in other areas. [25] Women remain underrepresented even though we observe the increased publishing rate for both genders during the pandemic. Similar disparity can be observed for the last authors as well as the solo-authored papers. The increased gender gap in publishing is persistent across the ten countries with the highest number of researchers.

For the papers on topics other than COVID-19, we do not observe this high discrepancy, and in the case of the *medRxiv* we observe more women than projected by the model. The overall gender disparity in research during the pandemic is mostly driven by the higher publication rate of papers on COVID-19 and related topics. It seems that such research is conducted disproportionately by men, as male authors are more likely to appear in first author positions on papers posted on preprint servers and published in peer-reviewed journals.

It appears that the most significant drop in proportion of female authors happened early into the pandemic. The proportion of women has been increasing gradually for some authorship categories. Note that the observed gradual increase is statistically significant but is very slow. One can think of a possible explanation for such a sudden drop and a subsequent gradual increase is that most of the COVID papers published early were various epidemic models focusing on cases and death counts. Many of the authors' affiliations were in departments of engineering, mathematics and physics that might have a different proportion of women than the population of scientists in biology and medicine. Since research in the biomedical field usually takes longer to conduct and publish, it can lead to a shift in the gender distribution later. However, this argument does not explain the phenomenon entirely, as the base gender gap in STEM is not higher than in biology [30] and the later publications from the biologists are not expected to narrow the gender gap. On the contrary, it might even increase it.

Another likely explanation of a sudden drop in the proportion of female authors is that caregiving demands have exploded during the pandemic, and these have mostly fallen on women. [12,34,35] These include childcare demands,[36] elderly care and other kinds of domestic and emotional labor. Sudden lockdowns and other preventive measures unevenly increased the burden on certain populations, causing the productivity of female scientists to decrease. As the world started fighting off the pandemic, people got used to the "new normal" and scientists started returning to their routines. That can partially explain the gradual increase in the proportion of female authors. Nevertheless, further research and more time is needed to investigate the reasons for such a sudden drop and gradual revival of the proportion of papers published by female scientists during the pandemic.

The global pandemic has touched almost every nation on the planet. Countries, however, responded differently to contain the spread of the disease. The variability of the measures and their timing, combined with differences in cultural norms and outbreak severity, have had a variable impact on researchers across the world. Country-level analysis better reveals global trends, as the aggregate data can be skewed by countries with a disproportionately large number of publications such as the United States, which has almost 29% of all authors in the data set (Multimedia Appendix 1, Table S10). Additionally, our analysis can reveal regional, political and cultural differences between the nations. It is known that gender disparities in research are strongly associated with a country's wealth.[37] The wealthier countries—with higher per capita gross domestic product (GDP)—have proportionally fewer female researchers, with Asian countries exhibiting the most pronounced gender disparity. However, the countries with higher GDP per capita are more resilient to the effects of COVID-19 on gender imbalance. In addition, wealthier countries show a smaller pandemic-related drop in women's participation in research than poorer countries, with wealthier Asian countries experiencing an increase in the proportion of active female researchers. This suggests that women experience bigger life disruptions in poorer countries, which affects their productivity. Additionally, women are more excluded from COVID-19 research in poorer nations. This certainly should not imply any purpose or deliberate action, but rather the disproportionate variations in the social environments across nations, caused by the various expectations for the female member of the household.

## Implications

Gender imbalances in academia have been evident historically and still persist today. Various measures of research output, including the proportion of authors, fractionalized authorships,[38] tenure decisions and number of research grants[39] indicate the significant gender gap that is observed worldwide. For the past 60 years, we have witnessed the increase of participation of women in science across scientific disciplines[30] and lower levels of discrimination,[40] demonstrating that the gender gap is shrinking over time.[41] Thus, a sudden drop in women's research output in biomedical research about COVID-19 appears as a surprising reverse trend.

The factors that lead to such extreme and consistent differences in the proportion of female scientists can be numerous. The already existing barriers for female participation in science vary across countries. In some nations, men are more favorably placed than women[42–45] and can be more likely to get quick funding for COVID-19 related research. Additionally, traditional gender norms differ and can affect the genders differently. Caregiving demands have mostly fallen on women. New challenges at the same time bring new opportunities, and men who are likely to engage in a more aggressive self-promotion[46,47] and pursue careers more forcefully[41] can be motivated to push for faster publication. Identifying the exact reasons for an increased gender gap can be an important topic for future studies.

The global pandemic caused this unforeseen crisis that will most certainly affect academia. All the difficulties female scientists faced previously may possibly be exacerbated by the extended lockdowns and sudden shift in work-life dynamics. It is important to understand the impact of such an extraordinary circumstance on the scientific community that will disproportionately affect research outputs as well as prospects for tenure and promotions.[48] Future research evaluation practices should be informed by our findings to account for and mitigate the penalizing effects that COVID-19 is having on female researchers.

## Strengths and limitations of study

The strengths of our study include the use of a relatively large and diverse dataset from three different publishing platforms. The focus on preprint papers allows the assessment of the observed effects in a timely manner. We focus on a structured and rigorous statistical analysis, making sure that the results are significant. The data and the code to reproduce the results are available.

Potential limitations warrant consideration. First, the gender of the author can be wrongly identified. Even though we excluded the results that have a low confidence, a small fraction of the authors can be misgendered. Additionally, we acknowledge that automated gender classifiers do not recognize the various non-binary gender identities[49] and assign the gender label based on the historical distribution of typical male/female names. As awareness of the nature of gender and identify shift, so may the number of researchers who do not identify within the binary categories of male and female. Such researchers face additional layers of discrimination that our study does not consider. While we understand that binary gender can be an oversimplification that can introduce some amount of bias and inaccuracy, the problem that we highlight hopefully can bring some attention to the multi-faceted issue of gender, identity and discrimination. Second, the algorithm that identifies the authors' country relies on recognizing the names of the toponyms in the name of the authors' affiliated institution. Even though we make sure that the most popular institutions are properly localized and we optimize the localization resolution, some errors are possible. Third, throughout the paper, the word "productivity" is used to refer to the rate of publication output of scientists in terms of publications per week and it cannot capture changes to scientists' other inputs. For example, female scientists appearing less productive in terms of publications-per-week may simply reflect that they are able to spend less time on their research, not capturing the publications-per-hour-worked. We are aware that there are other preprint servers and journals that publish papers in biomedical field. Analyzing the data from the two largest preprint servers and the largest publisher of peer-reviewed papers, we aim to cover a representative sample of papers and authors from the field. Finally, our analysis is focused on the first six months of the pandemic and might not accurately evaluate the effects that can be observed later in the pandemic.

## Conclusion

Our findings document a decrease in the proportion of female authors in biomedical field who published research papers during the global pandemic. This effect is particularly pronounced for papers related to COVID-19, indicating that women are producing fewer publications related to COVID-19 research. A sudden increase in this gender gap is persistent across the ten countries with the highest number of researchers. The results should be used to inform the scientific community of the worrying trend in COVID-19 research and the disproportionate effect the pandemic has on female academics' research outputs.

## Acknowledgements

The authors are grateful to the Defense Advanced Research Projects Agency (DARPA), contract W911NF192027, for their support.

All authors conceived and designed the study. GM collected and analyzed the data. All authors wrote and revised the manuscript.

## Conflicts of Interest

The authors have no relevant interests to declare.

## Abbreviations

OLS: Ordinary Least Squares
RDD: Regression Discontinuity Design